\documentclass{eas}

\usepackage{graphicx}

\usepackage[utf8]{inputenc}
\usepackage{xcolor}

\newcommand{\ind}[1]{_{\mathrm{#1}}}
\newcommand{\diff}{\mathrm{d}}
\newcommand\DPi{\Delta\Pi\ind{1}}
\newcommand\Dnu{\Delta\nu}

\newcommand\np{{n\ind{p}}}
\newcommand\ngrav{{n\ind{g}}}
\newcommand\nmax{n\ind{max}}
\newcommand\dl{d_{01}}
\newcommand\epsp{\varepsilon\ind{p}}
\newcommand\epsg{\varepsilon\ind{g}}
\newcommand\dnurot{\delta\nu\ind{rot}}
\newcommand\dnurotcore{\delta\nu\ind{rot, core}}

\newcommand\nup{\nu\ind{p}}
\newcommand\nug{\nu\ind{g}}

\newcommand{\BV}{Brunt-V\"ais\"al\"a}

\newcommand\Dtaum{\Delta\tau\ind{m}}
\newcommand\xrot{x\ind{rot}}
\newcommand\numax{\nu\ind{max}}

\begin{document}

\title{Measuring the core rotation of red giant stars} 
\author{Charlotte Gehan}\address{charlotte.gehan@obspm.fr}
\author{Beno\^it Mosser}
\author{Eric Michel}
\begin{abstract}
Red giant stars present mixed modes, which behave as pressure modes in the convective envelope and as gravity modes in the radiative interior. This mixed character allows to probe the physical conditions in their core. With the advent of long-duration time series from space-borne missions such as CoRoT and \textit{Kepler}, it becomes possible to study the red giant core rotation. As more than 15 000 red giant light curves have been recorded, it is crucial to develop a robust and efficient method to measure this rotation. Such measurements of thousands of mean core rotation would open the way to a deeper understanding of the physical mechanisms that are able to transport angular momentum from the core to the envelope in red giants.\\
In this work, we detail the principle of the method we developed to obtain automatic measurements of the red giant mean core rotation. This method is based on the stretching of the oscillation spectra and on the use of the so-called Hough transform. We finally validate this method for stars on the red giant branch, where overlapping rotational splittings and mixed-mode spacings produce complicated frequency spectra.
\end{abstract}
\maketitle
\section{Introduction}

Red giant stars present mixed modes, which behave as pressure modes in the convective envelope and as gravity modes in the radiative interior (Beck et al. \cite{Beck}). This mixed character allows to study the core of red giants. This is not the case for main-sequence stars, where gravity modes are confined in the inner radiative zone and evanescent in the convective envelope. Red giants differ from main sequence stars because they have a very dense radiative core, where the \BV\ frequency reaches high values (Montalb\'an et al. \cite{Montalban}). Moreover, the evanescent region between gravity and pressure mode resonant cavities is much narrower than in the case of main-sequence stars (Goupil et al. \cite{Goupil}). Thus physical conditions in red giants are met to cause a coupling between pressure modes in the convective envelope and gravity modes in the radiative interior, giving birth to non-radial mixed modes.\\
The study of red giant oscillation spectra led to the automatic measurement of many seismic parameters such as the large separation $\Dnu$ (Mosser et al. \cite{Mosser_2009}), the frequency of maximum power oscillation $\numax$ (Kallinger et al. \cite{Kallinger}), and the gravity mode period spacing $\DPi$ (Vrard et al. \cite{Vrard}). Many information can be retrieved from these parameters. The measurement of $\Dnu$ and $\numax$ allows to determine stellar masses and radii with a precision between 4 to 8 \% for the radius and 8 to 16 \% for the mass (Kallinger et al. \cite{Kallinger}). The measurement of $\DPi$ provides the determination of stars evolutionary status: we can now distinguish between the subgiant phase, the red giant branch (RGB), the red clump where stars burn their helium in their core, and the beginning of the asymptotic giant branch (Mosser et al. \cite{Mosser_2014}).\\
Although the measurement of all these parameters is now entirely automated, this is not the case for the measurement of rotational splittings.
The difficulty comes from the fact that mixed modes are not evenly spaced in frequency: pressure-dominated mixed modes are nearly equally spaced in frequency with a spacing close to the large separation $\Dnu$, while gravity-dominated mixed modes are nearly equally spaced in period with a spacing close to the gravity mode period spacing $\DPi$ (Mosser et al. \cite{Mosser_2012b}). Red giant spectra exhibit numerous dipole mixed modes, which apparently form an inextricable mixed mode forest. The asymptotic expansion of mixed modes is used to identify the mixed mode pattern. In the case of rapid core rotation, it is however difficult to disentangle the frequency spacings between two consecutive mixed modes from the rotational splittings in frequency oscillation spectra.\\ Measurements of the mean core rotation were manually obtained for about 300 red giants (Mosser et al. \cite{Mosser_2012c}). They showed that a very efficient angular momentum transport from the core to the envelope is at work in red giants. The physical mechanism responsible for this transport is still not yet fully understood. Mean core rotation measurements for a red giant set as large as possible are therefore needed to constrain the nature of this mechanism.\\
In this work, we detail the principle of the method developed in order to obtain automated measurements of the mean core rotation of red giants. We aim at validating it for stars on the RGB, for which automated measurements are fully consistent with manual measurements (Mosser et al. \cite{Mosser_2012c}). Such automated measurements are essential to pave the way for the future analysis of PLATO mission data, with a potential as high as half a million red giants.

\section{Stretching the spectra}

The first step of the method consists in stretching the frequency spectra, which provides spectra where mixed modes are now regularly spaced with a spacing close to $\DPi$. Therefore, frequencies are changed into stretched periods $\tau$ through the differential equation (Mosser et al. \cite{Mosser_2015})
\begin{equation}
\diff \tau = \frac{\diff \nu}{\zeta \nu^2}.
\end{equation}
The $\zeta$ function is defined by
\begin{equation}
\zeta = \left[1 + \frac{1}{q} \frac{\nu^2 \DPi}{\Dnu} \frac{\cos^2 \left[\pi \frac{1}{\DPi} \displaystyle{\left(\frac{1}{\nu} - \frac{1}{\nug} \right)}\right]}{\cos^2 \displaystyle{\left(\pi \frac{\nu - \nup}{\Dnu}\right)}} \right]^{-1},
\end{equation}
where $q$ is the coupling parameter of mixed modes, $\DPi$ is the gravity mode period spacing, $\Dnu$ is the large separation, $\nug$ are the pure dipole gravity mode frequencies, $\nup$ are the pure pressure mode frequencies.\\
For the pure dipole gravity mode frequencies $\nug$ we can use the first-order asymptotic expansion (Tassoul \cite{Tassoul})
\begin{equation}
\frac{1}{\nug} = -(\ngrav + \epsg) \, \DPi,
\end{equation}
where $\ngrav$ is the gravity radial order usually defined as a negative value, and $\epsg$ is a small but complicated function sensitive to the stratification near the boundary between the radiative
core and the convective envelope.\\
For the pure dipole pressure mode frequencies $\nup$ we use the universal red giant oscillation pattern (Mosser et al. \cite{Mosser_2012b})
\begin{equation}
\nup = \left(\np + \frac{1}{2} + \epsp + \dl + \frac{\alpha}{2} [\np - \nmax]^2 \right) \Dnu,
\end{equation}
where $\np$ is the pressure radial order, $\epsp$ is the pressure mode offset, $\dl$ is the small separation, $\alpha$ represents the curvature of the oscillation pattern, and $\nmax = \numax / \Dnu - \epsp$ is the non-integer order at the frequency $\numax$ of maximum oscillation signal.



\section{Rotation signature in stretched spectra}

\begin{figure}[t]
\centering
\includegraphics[width=9.5cm]{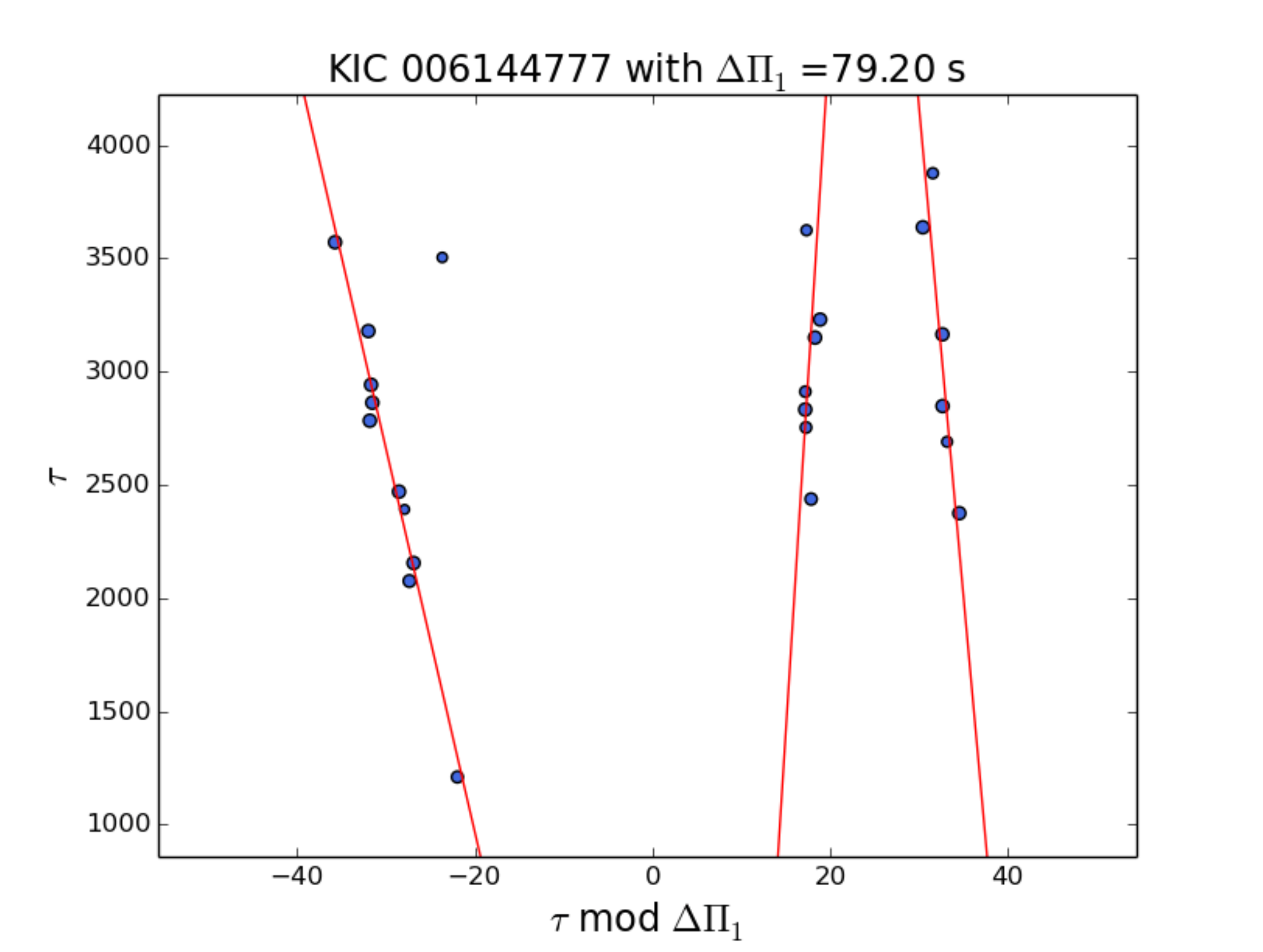}
\caption{Échelle diagram representing $\tau$ as a function of $\tau$ modulo $\DPi$ for the star KIC 6144777. The rotational multiplet components corresponding to the azimuthal orders $m=\lbrace -1, 0, 1\rbrace$ are represented by blue dots. Red lines represent the automatic identification of each component through the Hough transform.}
\label{fig-1}
\end{figure}

In the stretched spectra, rotation induces a small departure from an evenly spaced pattern. The stretched period spacing between rotational multiplet components with the same azimuthal order $m$ expresses (Mosser et al. \cite{Mosser_2015})
\begin{equation}
\Dtaum = \DPi (1 + m \xrot).
\label{Dtaum}
\end{equation}
It presents a small departure from $\DPi$, expressed by
\begin{equation}
\xrot = 2 \, \zeta \, \frac{\dnurotcore}{\nu},
\label{xrot}
\end{equation}
where $\dnurotcore$ is the rotational splitting induced by the mean rotation of the radiative core.\\
In the case of rapid core rotation, this equation can be used to characterize the crossing of the multiplet components and to infer $\dnurotcore$ (Gehan et al. \cite{Gehan}). When $\xrot \ll 1$ for slow rotation, a relevant approximation of $\xrot$ is given by (Mosser et al. \cite{Mosser_2015})
\begin{equation}
\xrot \simeq 2 \, \frac{N}{N+1} \frac{\dnurotcore}{\numax},
\end{equation}
 where $N$ is the number of gravity modes per $\Dnu$-wide frequency range, defined as
\begin{equation}
N = \frac{\Dnu}{\DPi \, \numax^2}.
\label{N}
\end{equation}
It is then possible to build échelle diagrams based on this stretched period $\tau$, where the different components of the rotational multiplet draw ridges and are disentangled (Figure~\ref{fig-1}). Hence, it is easy to obtain measurements of the mean core rotation. The number of ridges that are visible depends on the stellar inclination. Only the ridge associated to the azimuthal order $m = 0$ is visible if the star is seen pole-on, whereas the two components $m = \pm \, 1$ are visible when the star is seen equator-on, and finally all the three components $m = \lbrace -1, 0, 1 \rbrace$ are visible for intermediate inclinations.\\
In the case of rapid core rotation, the ridges overlap in the échelle diagram, and the frequency where the crossing occurs together with the knowledge of the $\DPi$ parameter provide a precise estimate of the mean core rotation value (Gehan et al. \cite{Gehan}).

\section{Measuring rotational splittings: the Hough transform}

\begin{figure}[t]
\centering
\includegraphics[width=\hsize,clip]{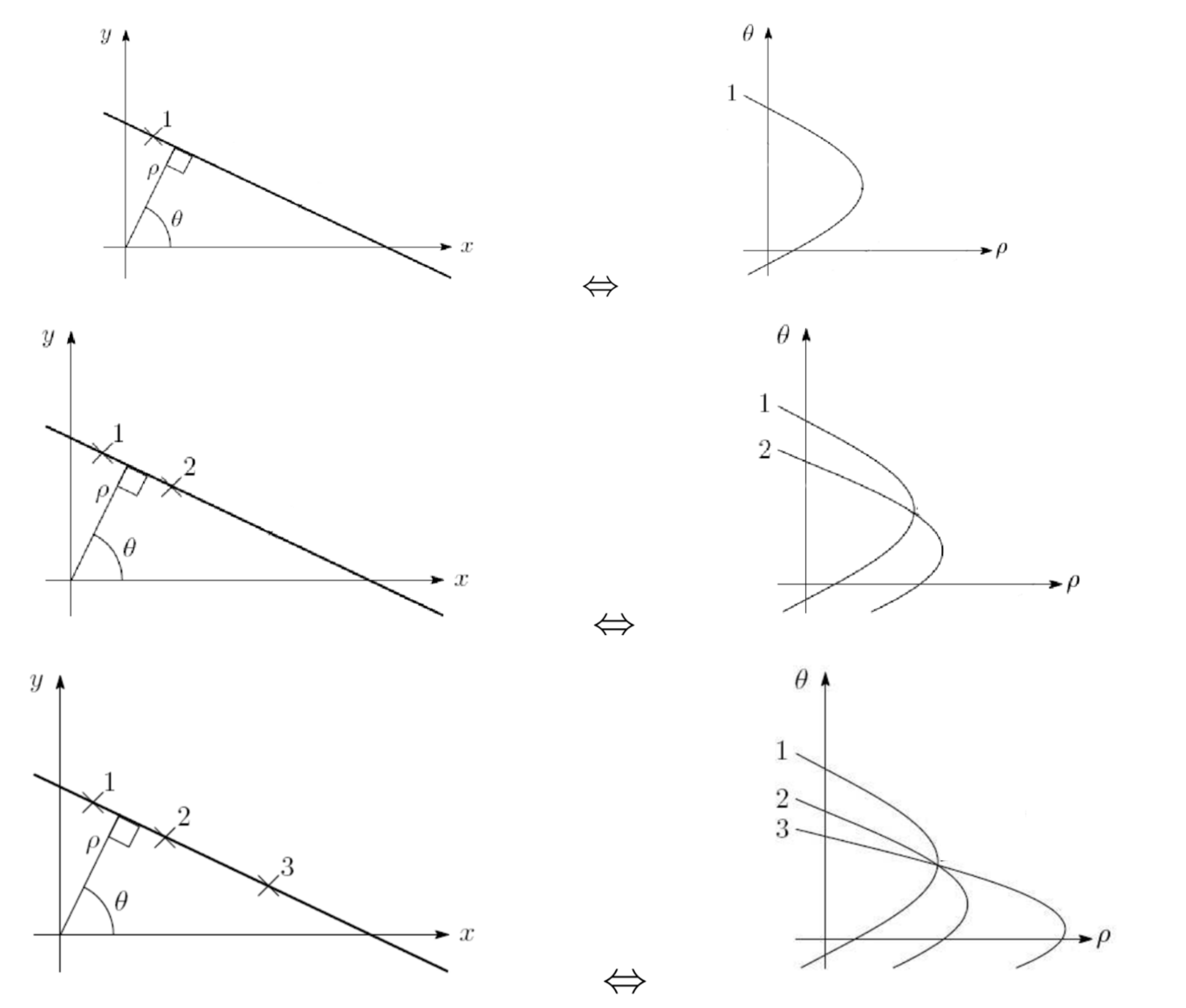}
\caption{Principle of the Hough transform (\texttt{http://slideplayer.fr/slide/1567709/}). A point in the ($x,y$) space (top left) is represented by a sinusoid in the Hough space ($\rho$, $\theta$) (top right). Points which are aligned in the ($x,y$) space (middle and bottom left) are represented by sinusoids which all intersect at the same ($\rho$, $\theta$) values in the Hough space (middle and bottom right). Thus a line is characterized by a couple of ($\rho$, $\theta$) values.}
\label{fig-2}
\end{figure}

We aim at identifying the ridges corresponding to the different rotational multiplet components (Figure~\ref{fig-1}). According to Equation~\ref{Dtaum}, these ridges are nearly vertical in a stretched échelle diagram. The measurement of the rotation can be derived from the identification of these lines, in the form of $y=ax+b$, where $x$ and $y$ are the cartesian coordinates of the points, $a$ is the slope and $b$ is the intercept of the line. Detecting lines using the previous equation  proves to be problematic for lines which are almost vertical. In such cases, the slope tends towards infinity and the determination of the slope and the intercept is highly imprecise. The Hough transform\footnote{The classical Hough transform is a feature extraction technique commonly used to identify lines in an image, but it can also be used to identify arbitrary shapes as circles or ellipses (Ballester \cite{Ballester}).} allows to get around this problem through the use of the polar coordinates ($\rho$, $\theta$) that characterize the position and orientation of the line (Figure~\ref{fig-2}): $\theta$ corresponds to the angle between the line which is perpendicular to the alignment that we consider and the x-axis, $\rho$ represents the distance between the alignment that we consider and the origin.
The Hough transform is particularly useful to detect the different ridges drawn by red giant core rotation in échelle diagrams, which are quasi vertical.\\
The relation between $\rho$ and $\theta$ is:
\begin{equation}
\rho = x \cos \theta + y \sin \theta.
\label{tho}
\end{equation}
For an isolated point, the $\theta$ angle may take all possible values between $-\pi/2$ and $\pi/2$. A point have fixed ($x$, $y$) values and is thus represented by a sinusoid in the Hough space ($\rho$, $\theta$) (Figure~\ref{fig-2} top). If several points are aligned on the same line, the corresponding sinusoids all intersect at the same ($\rho$, $\theta$) coordinates in the Hough space (Figure~\ref{fig-2} middle and bottom). Therefore, detecting lines in the ($x,y$) space is equivalent to detect intersections in the Hough space.
In practice, the algorithm looks for points which are almost aligned and almost vertical in échelle diagrams to identify in an automated way the different ridges.\\
From the Hough parameters, we can derive the mean core rotation, since the angle $\theta$ used in the Hough transform is linked to $\xrot$ by
\begin{equation}
\tan \theta = m \xrot.
\label{tan_theta}
\end{equation}
The correction to the $\DPi$ gravity mode spacing induced by rotation in stretched spectra is small. In practice, $\xrot \leq 0.02$, so that $\theta \leq 0.01$. Hence, we have the direct relation
 \begin{equation}
\theta \simeq  2 \, m\, \frac{N}{N+1} \frac{\dnurotcore}{\numax},
\label{theta}
\end{equation}
from which we can derive $\dnurotcore$.


\begin{figure}[t]
\centering
\includegraphics[width=\hsize,clip]{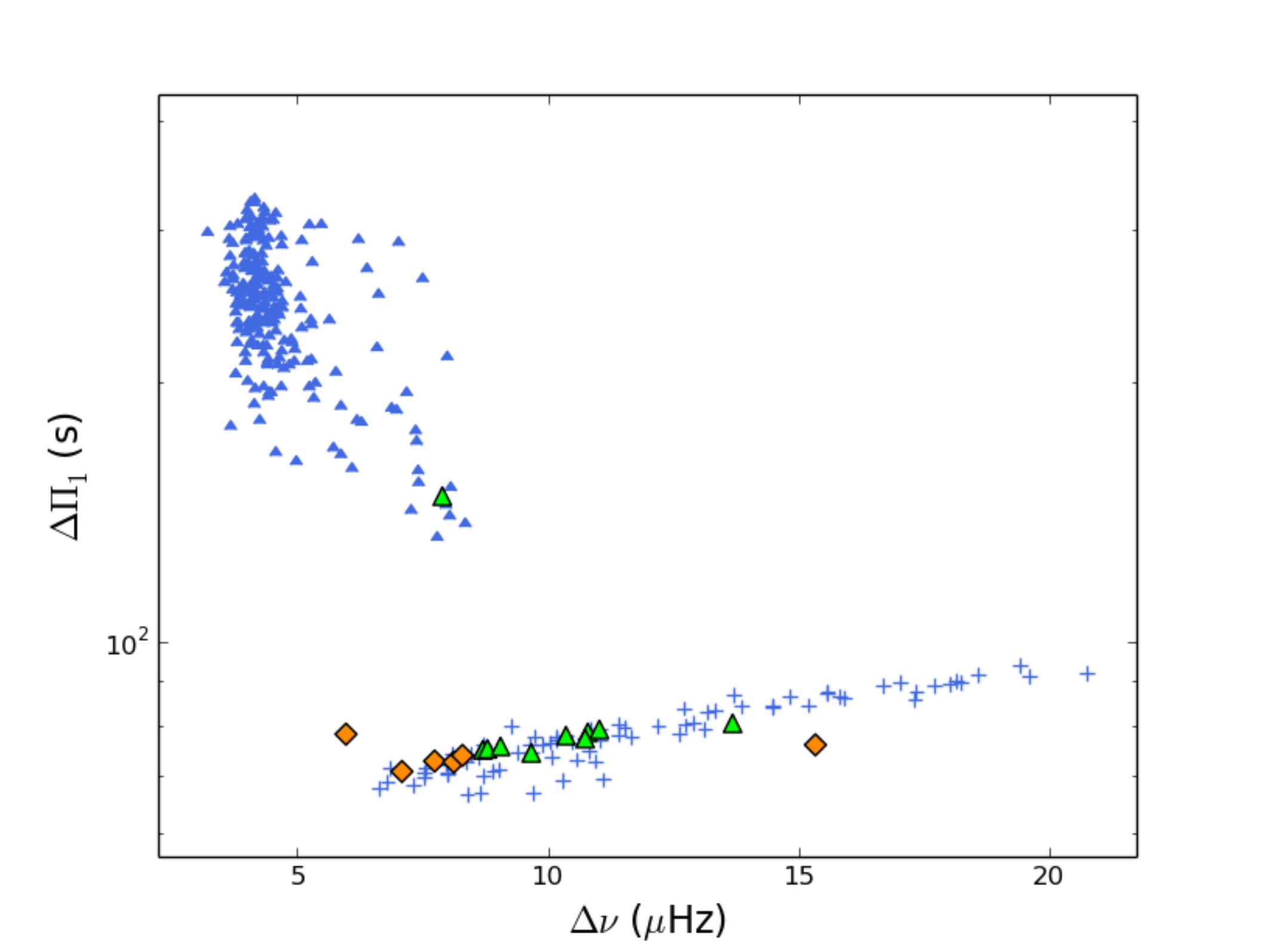}
\caption{$\DPi$-$\Dnu$ diagram. Blue symbols represent the sample analysed by Mosser et al. \cite{Mosser_2012c}, with blue crosses accounting for RGB stars and blue triangles accounting for red clump stars. Green triangles represent the stars on which we applied the automatic method developed in this study, orange diamonds represent the rapidly rotating red giants analysed by Gehan et al. \cite{Gehan}.}
\label{fig-3}
\end{figure}

\section{Validation of the method on RGB stars}

We applied this method to eight red giants. The obtained measurements are resumed in Table 1. With the use of the Hough transform, we could obtain core rotational splitting measurements with a relative uncertainty on the order of a few percents. The sample represents seven RGB stars and a red giant which is transiting to the clump phase (Figure~\ref{fig-3}). RGB stars are located very close to the confusion limit in frequency between the rotational splitting $\dnurot$ and the spacing between consecutive mixed modes (Mosser et al. \cite{Mosser_2012c}). That means that the rotational splitting and the mixed-mode frequency spacing sometimes overlap for RGB stars. Thus measuring the mean core rotation in RGB stars is particularly difficult. Nevertheless, we could obtain automatic measurements for stars in this evolutionary stage. In these conditons, obtaining automatic measurements of the mean core rotation of red clump stars will be easier providing that we develop a method which would be more appropriate to their lower core rotation values.

\begin{table}
\centering
\caption{Core rotational splittings  $\dnurotcore$, uncertainties $\delta (\dnurotcore)$ and relative uncertainties $\delta (\dnurotcore) / \dnurotcore$ measured with the Hough transform}
\label{tab-1}       
\begin{tabular}{cccccc}
\hline
KIC & $\DPi$ & $\Dnu$ & $\dnurotcore$ & $\delta (\dnurotcore)$ & $\delta (\dnurotcore) / \dnurotcore$\\
 & \small{(s)} & \small{($\mu$Hz)} & \small{(nHz)} & \small{(nHz)} & \small{(\%)}\\\hline
3526061 & 77.3 & 10.741 & 412 & 18 & 4.3\\
6144777 & 79.2 & 11.025 & 266 & 31 & 11.6\\
7430666 & 78.6 & 10.796 & 287 & 36 & 12.6\\
8081497 & 75.7 & 9.054 & 597 & 20 & 3.4\\
8475025 & 74.4 & 9.667 & 274 & 10 & 3.6\\
9267654 & 77.9 & 10.359 & 497 & 15 & 3.0\\
10866415 & 75.2 & 8.794 & 323 & 41 & 12.6\\
11550492 & 75.0 & 8.699 & 378 & 37 & 9.8\\\hline
\end{tabular}
\end{table}





\section{Conclusions}

Disentangling rotational splittings from mixed modes is now possible with the use of stretched spectra. We developed a largely automated method to measure the mean core rotation and validated it on RGB stars, where large rotational splittings lead to complicated frequency oscillation spectra.  The entire automation of red giant core rotation measurement is in progress, and we aim at obtaining mean core rotation measurements for thousands of red giants observed by \textit{Kepler} in the near future. These measurements will be essential to pave the way for the future analysis of PLATO data, with potentially 500 000 red giants. Such measurements will also allow us to get more information on the physical mechanisms responsible for angular momentum transport in these stars, thus improving our understanding of stellar physics in deep stellar interiors.



\end{document}